\documentclass[10pt]{article}
\usepackage{amsmath}
\usepackage{amssymb}
\usepackage{natbib}

\begin{document}
\bibliographystyle{plainnat}
\setcitestyle{numbers,square}

\title{Gravitational quantization of exoplanet orbits                    \\
                   in 55~Cnc, $\upsilon$~And, Kepler-11, Kepler-20,      \\
                   and Kepler-90}

\author{Vassilis S. Geroyannis \\
        Department of Physics, University of Patras, Greece          \\  
        vgeroyan@upatras.gr}

\maketitle

\begin{abstract}
In the framework of the so-called ``global polytropic model'', we assume hydrostatic equilibrium for a planetary system, and solve the resulting Lane--Emden differential equation in the complex plane. We thus obtain polytropic spherical shells defined by succesive roots of the real part $\mathrm{Re}(\theta)$ of the Lane-Emden function $\theta$. These shells seem to provide hosting orbits for the planets of the system(s) under consideration. In the present investigation, we study within this framework the exoplanet systems 55~Cnc, $\upsilon$~And, Kepler-11, Kepler-20, and Kepler-90.   
\end{abstract}

\section{Introduction}
\label{intro}
Gravitational quantization of orbits in systems of planets or satellites has been considered in two recent investigations (\citep{GVD14}, \citep{G14}). The equations of hydrostatic equilibrium, assumed to govern such systems in the framework of classical mechanics, 
yield the well-known Lane--Emden differential equation, which is solved in the complex plane according to the so-called ``complex plane strategy'' (details on this issue are given in \citep{GKAR14}); the solution obtained is the complex Lane--Emden function $\theta$.  Polytropic spherical shells defined by succesive roots of its real part $\mathrm{Re}(\theta)$ seem to be appropriate places for accomodating planets or satellites of the system(s) considered.
As emphasized in \citep{G14}, there is only one parameter to be adjusted for a particular polytropic configuration defined by $\theta$: the polytropic index $n$ of the central body.
  
Alternative studies regarding quantized orbits of planets or satellites are discussed and cited in \citep{GVD14} and \citep{G14}.

In this study, we apply the global polytropic model to the exoplanet systems 55~Cnc, $\upsilon$~And, Kepler-11, Kepler-20, and Kepler-90.

\section{Polytropic Shells Hosting Planets}
\label{gpm}
The complex-plane strategy and the complex Lane--Emden function $\theta$ are analyzed in \citep{GKAR14} (Sec.~3.1). Preliminary concepts regarding the global polytropic model are presented in \citep{GVD14} (Sec.~3). For convenience, we will use hereafter the definitions and symbols adopted in \cite{GVD14}. 

The real part $\bar{\theta}(\xi)$ of the complex function $\theta(\xi)$ has a first root at $\xi_1 = \bar{\xi_1} + i \, \breve{\xi_0}$, a second root at $\xi_2 = \bar{\xi_2} + i \, \breve{\xi_0}$ with $\bar{\xi_2} > \bar{\xi_1}$, a third root at $\xi_3 = \bar{\xi_3} + i \, \breve{\xi_0}$ with $\bar{\xi_3} > \bar{\xi_2}$, etc. The polytropic sphere of polytropic index $n$ and radius $\bar{\xi}_1$ is the central component of a resultant polytropic configuration with further components the polytropic spherical shells $S_2$, $S_3$, \dots, defined by the pairs of radii $(\bar{\xi}_1, \, \bar{\xi}_2)$, $(\bar{\xi}_2, \, \bar{\xi}_3)$, \dots, respectively.
Each polytropic shell can be considered as an ideal hosting place for a planet or satellite. The most appropriate orbit radius $\alpha_j \in [\bar{\xi}_{j-1},\,\bar{\xi_j}]$ is that at which $|\bar{\theta}|$ takes its maximum value inside $S_j$, 
\begin{equation}
\mathrm{max}|\bar{\theta}[S_j]| = |\bar{\theta}(\alpha_j + i \, \breve{\xi}_0)|.
\label{maxth} 
\end{equation}

In the case of two planets or satellites hosted inside the same shell $S_j$, there are two proper orbits with radii $\alpha_\mathrm{Lj}$ and $\alpha_\mathrm{Rj}$, such that $\alpha_\mathrm{Lj} < \alpha_j < \alpha_\mathrm{Rj}$, at which $|\bar{\theta}|$ becomes equal to its average value inside $S_j$,
\begin{equation}
\mathrm{avg}|\bar{\theta}[S_j]| = |\bar{\theta}(\alpha_\mathrm{Lj} + i \, \breve{\xi}_0)|
                                = |\bar{\theta}(\alpha_\mathrm{Rj} + i \, \breve{\xi}_0)|.
\label{avgth}   
\end{equation}
Accordingly, two planets or satellites inside $S_j$ can be hosted on orbits with radii  $\alpha_\mathrm{Lj}$ and $\alpha_\mathrm{j}$, or, alternatively,  $\alpha_\mathrm{j}$ and $\alpha_\mathrm{Rj}$.   

An algorithm for computing the optimum polytropic index $n_\mathrm{opt}$ of a star with a system of planets, or of a planet with a system of satellites, is presented in \citep{G14} (Sec.~2). This algorithm, so-called A[n], can be applied to $N_\mathrm{P}$ members $P_1,\,P_2,\,\dots,\,P_{N_\mathrm{P}}$ of such a system with $N_\mathrm{P}$ prescribed distances $A_{1} < A_{2} < \dots < A_{{N_\mathrm{P}}}$ from the central body.

\section{Polytropic Models Simulating Host Stars and Computations}  
The host stars of the exoplanet systems 55~Cnc (\citep{BBB11}, \citep{ERC12}, \citep{NFW14}) $\upsilon$~And (\citep{AF99}, \citep{CGCP11}, \citep{DBM14}), Kepler-11 (\citep{MW14}, \citep{BMN14}), Kepler-20 (\citep{FTR11}, \citep{GCR12}), and Kepler-90 (\citep{CCL13}, \citep{SDF14}) are Sun-like stars. It is therefore expected that appropriate values of the polytropic index $n$ for modelling such stars are about $n \sim 3$ (see e.g. \citep{Hor04}, Sec.~6.1 and references therein; see also \citep{GVD14}, Sec~3 and references therein). 
In this study, we apply the general algorithm A[n] to an array $\{n_i\}$ with elements
\begin{equation}
n_i = 2.500 + 0.001 \, (i-1), \qquad i = 1, \, 2, \, \dots, \, 1001.
\label{Nn-now}
\end{equation} 
Accordingly, we expect to find optimum values of the polytropic index for the Sun-like stars of our study in the interval 
\begin{equation}
\mathbb{I}_n = [2.500, \, 3.500]. 
\label{In}
\end{equation}
The 1001 complex IVPs counted in Eq.~\ref{Nn-now} are solved by \texttt{DCRKF54} \cite{GV12}, which is a Runge--Kutta--Fehlberg code of fourth and fifth order modified for solving complex IVPs of high complexity in their ODEs, along contours prescribed as continuous chains of straight-line segments (details on the usage of \texttt{DCRKF54} are given in \citep{GVD14}, Sec.~4). Integrations proceed along the contour  
\begin{equation}
\mathfrak{C} = \{\xi_0 = (10^{-4},\,10^{-4}) \rightarrow 
                               \xi_\mathrm{end} = (1.0 \times 10^5,\,10^{-4})
                               \};
\label{CJSUN}
\end{equation}
this contour is of the special form~(8) of \citep{GVD14} (various contours and their characteristics are defined in \cite{GV12}, Sec.~5).

\section{Numerical Results and Discussion}
\label{results}
Since physical interest focuses on real parts of complex orbit radii, we will hereafter quote only such values and, for simplicity, we will drop overbars denoting such real parts. 

For convenience, all astrophysical data used for comparing our results with respective exoplanet observations are those appearing in http://exoplanet.eu. The first root $\xi_1$ of $\theta$, coinciding with the radius of the host star, is expressed in both ``classical polytropic units'' (cpu) --- in these units, the length unit is equal to the polytropic parameter $\alpha$ (\citep{GVD14}, Eq.~(3b)) --- and solar radii $R_\odot$. All other orbit radii are expressed in AU.    

Numerical results computed by A[n] are given in Tables~\ref{55Cnc}--\ref{k90IE}. In detail, Table~\ref{55Cnc} shows results for the 55~Cnc exoplanet system; the minimum sum of absolute percent errors is 
\begin{equation}
\begin{aligned}
\Delta_\mathrm{min} 
\biggl( n_\mathrm{opt}(\mathrm{55\,Cnc}) & =
        3.125; \, q_\mathrm{e} = 2, \biggr. \\
      & \biggl. q_\mathrm{b} = 4, \, 
                q_\mathrm{c} = 5, \, q_\mathrm{f} = 7, \, q_\mathrm{d} = 13 
\biggr) \simeq 40.9;
\label{DminJnow}
\end{aligned}
\end{equation}
the symbols are from this table. Smaller error is that for d's distance, $\simeq 0.19\%$; d is also the most massive planet in the 55~Cnc system. Larger error is that for e's distance, $\simeq 32\%$. It is worth remarking here that this error is the larger one among all systems examined in the present investigation. It may be so due to the proximity of the shell accomodating planet e (shell No~2) with the host star. In fact, 55 Cnc is the only system, examined here, with a planet hosted in the innermost shell.
 
The average error for the computed orbit radii of the five planets in 55~Cnc is $\sim 8\%$.

Next, Table \ref{55CncIE} gives results for all internal shells, i.e. shells located inside the last occupied shell, which seem to be unoccupied according to the up-to-now observations. On this matter, we find interesting to compare orbit radii of planets predicted in \citep{BL13} (Table~2) with radii computed here. Such comparisons are also made for external shells, i.e. shells located outside the last occupied shell, with predicted planets. In particular, the internal shells No~3 and No~9 provide hosting orbits with radii differing $\sim 11\%$ and $\sim 13\%$ relative to the distances predicted in \citep{BL13}. In addition, the external shell No~17 provides a planet orbit with radius differing only $\sim 0.5\%$ relative to that predicted by \citep{BL13}.       

The optimum case for the $\upsilon$~And system has minimum sum of absolute percent errors (results and symbols are from Table~\ref{uAn})
\begin{equation}
\Delta_\mathrm{min} \biggl( n_\mathrm{opt}(\upsilon\,\mathrm{And}) = 2.949; \, q_\mathrm{b} = 3, \, q_\mathrm{c} = 8, \, q_\mathrm{d} = 11, \, q_\mathrm{e} = 14 \biggr) \simeq 12.5.
\label{DminS}
\end{equation}
Here, smaller error is that for e's distance, $\simeq 0.24\%$, and larger one that for b's distance, $\simeq 7.5\%$. For the most massive planet, d, the corresponding error is $\simeq 4.5\%$. 

Accordingly, the average error for the computed distances of the four planets in $\upsilon$~And is $\sim 3\%$.

Furthermore, Table \ref{uAnIE} shows that the internal shells No~4 and No~5 provide hosting orbits, of which the radii exhibit discrepancies $\sim 12\%$ and $\sim 14\%$, respectively, relative to the radii predicted by \citep{BL13}. In addition, the external shell No~20 hosts an orbit with radius differing $\sim 0.5\%$ relative to that predicted in \citep{BL13}. 
 
In the case of the Kepler-11 system of exoplanets, Table~\ref{k11} gives for its optimum case (we use the symbols of this table)  
\begin{equation}
\begin{aligned}
\Delta_\mathrm{min} \biggl( n_\mathrm{opt}(\mathrm{Kepler\!-\!11}) & = 
2.779; \, q_\mathrm{b}=5, \biggr. \\ 
& \biggl. 
  q_\mathrm{c}=5, \, q_\mathrm{d}=6, \, q_\mathrm{e}=6, \, q_\mathrm{f}=7, \, 
  q_\mathrm{g}=9 \biggr) \simeq 32.7.
\label{DminU}
\end{aligned}
\end{equation}
Smaller error is that for g's distance, $\simeq 2.3\%$, which is also the most massive planet of this system. Larger error is that for e's distance, $\simeq 8.7\%$;

It is worth remarking that the shell No~5 is occupied by the two planets b and c. The former is resident of the ``maximum-density orbit'' (Eq.~(\ref{maxth})) with radius $\alpha_\mathrm{b} = \alpha_5$; and the latter is hosted on the ``average-density orbit'' (Eq.~(\ref{avgth})) with radius $\alpha_\mathrm{c} = \alpha_\mathrm{R5}$. Likewise, the shell No~6 is occupied by the planets d and e. The former is resident of the average-density orbit with radius $\alpha_\mathrm{d} = \alpha_\mathrm{L6}$; while the latter is resident of the maximum-density orbit with radius $\alpha_\mathrm{e} = \alpha_\mathrm{6}$. 

The average error in the computed distances of the six planets in Kepler-11 is $\simeq 5.5\%$.

Table \ref{k11IE} reveals that the internal shell No~8 provides a hosting orbit, of which the radius differs $\sim 8.5\%$ relative to the radius predicted in \citep{BL13}. The external shell No~20 does also host an orbit with radius differing $\sim 4.5\%$ relative to that predicted in \citep{BL13}. 

Regarding the planetary system of Kepler-20, Table~\ref{k20} gives the following optimum case (the symbols are from this table)  
\begin{equation}
\begin{aligned}
\Delta_\mathrm{min} \biggl( n_\mathrm{opt}(\mathrm{Kepler\!-\!20}) & = 
2.799; \, q_\mathrm{b}=4, \biggr. \\ 
& \biggl. 
  q_\mathrm{e}=4, \, q_\mathrm{c}=5, \, q_\mathrm{f}=5, \, q_\mathrm{d}=8
  \biggr) \simeq 19.5.
\label{DminK20}
\end{aligned}
\end{equation}
Smaller error is that for d's distance, $\simeq 0.1\%$, which is also the most massive planet in this system. Larger error is that for c's distance, $\simeq 10\%$.

Here, Table \ref{k20} shows that the shell No~4 is occupied by the two planets b and e. The former is resident of the average-density orbit with radius $\alpha_\mathrm{b} = \alpha_\mathrm{L4}$; while the latter is resident of the maximum-density orbit with radius $\alpha_\mathrm{e} = \alpha_\mathrm{4}$. Likewise, the shell No~5 is occupied by the planets c and f. The former is resident of the maximum-density orbit with radius $\alpha_\mathrm{c} = \alpha_5$; the latter is hosted on the average-density orbit with radius $\alpha_\mathrm{f} = \alpha_\mathrm{R5}$.  

The average error in the computed distances of the five planets in Kepler-20 is $\sim 4\%$.

Concerning unoccupied shells, Table \ref{k20IE} shows that the internal shell No~7 provides a hosting orbit, of which the radius differs $\sim 14.5\%$ relative to the radius predicted in \citep{BL13}. In addition, the external shell No~10 hosts an orbit with radius differing $\sim 16\%$ relative to that predicted in \citep{BL13}. 

Finally, for the planetary system of Kepler-90, Table~\ref{k90} gives the optimum case (the symbols are from this table)  
\begin{equation}
\begin{aligned}
\Delta_\mathrm{min} \biggl( n_\mathrm{opt}(\mathrm{Kepler\!-\!90}) & = 
2.819; \, q_\mathrm{b}=4, \, q_\mathrm{c}=4, \biggr. \\ 
& \biggl. 
    \, q_\mathrm{d}=7, \, q_\mathrm{e}=8, \, q_\mathrm{f}=8, \,
    q_\mathrm{g}=9, \, q_\mathrm{h}=11
  \biggr) \simeq 13.4.
\label{DminK90}
\end{aligned}
\end{equation}
Here, smaller error is that for b's distance, $\simeq 0.15\%$, while larger one is that for e's distance, $\simeq 5\%$. We also remark that the shell No~4 hosts the two planets b and c. Likewise, the shell No~8 hosts the planets e and f. This situation is similar to that appearing in the planetary system Kepler-11.  

The average error in the computed distances of the seven planets in Kepler-90 is $\sim 2\%$.

Finally, Table \ref{k90IE} shows all unoccupied internal shells and the lower three external shells. There are not predictions made in \citep{BL13} for this system.

\begin{table}
\begin{center}
\caption{The 55 Cnc system: central body $S_1$, i.e. the host star 55 Cnc, and polytropic spherical shells of the planets e, b, c, f, d. For successive shells $S_j$ and $S_{j+1}$, inner radius of $S_{j+1}$ is the outer radius of $S_j$. All radii are expressed in AU, except for the host's radius $\xi_1$. Percent errors of computed orbit radii $\alpha$ are given with respect to corresponding observed radii $A$, $100 \times |(A - \alpha)| / A$. Parenthesized signed integers following numerical values denote powers of 10. \label{55Cnc}}
\begin{tabular}{lr} 
\hline \hline
Host star 55 Cnc -- Shell No                    & 1              \\
$n_\mathrm{opt}$                                & $3.1250(+00)$   \\
$\xi_1$ (cpu)                                   & $7.4186(+00)$  \\
$\xi_1$ ($R_\odot$)                             & $9.4300(-01)$  \\
\hline

e -- Shell No                                   & 2                 \\
Inner radius, $\, \xi_1$                        & $4.3871(-03)$     \\
Outer radius, $\xi_2$                           & $2.5224(-02)$     \\
Orbit radius, $\, \alpha_\mathrm{e}=\alpha_2$   & $1.0645(-02)$     \\
Percent error in $\alpha_\mathrm{e}$, given that $A_\mathrm{e} = 1.56(-02)$   
& $3.18(+01)$              \\ 
\hline

b -- Shell No                                                  & 4                 \\
Inner radius, $\, \xi_3$      & $7.7733(-02)$                    \\
Outer radius, $\xi_4$                                          &  $1.7812(-01)$    \\
Orbit radius, $\, \alpha_\mathrm{b}=\alpha_4$  & $1.1650(-01)$     \\
Percent error in $\alpha_\mathrm{b}$, given that $A_\mathrm{b} = 1.134(-01)$   
& $2.73(+00)$              \\ 
\hline

c -- Shell No                                           & 5               \\
Outer radius, $\xi_5$                                   & $3.4540(-01)$     \\
Orbit radius, $\, \alpha_\mathrm{c}=\alpha_5$                              & $2.4719(-01)$     \\
Percent error in $\alpha_\mathrm{c}$, given that $A_\mathrm{c} = 2.403(-01)$     
& $2.87(+00)$     \\ 
\hline

f -- Shell No                                                 & 7               \\
Inner radius, $\, \xi_6$              & $5.9715(-01)$     \\
Outer radius, $\xi_7$              & $9.5598(-01)$     \\
Orbit radius, $\, \alpha_\mathrm{f}=\alpha_7$         & $7.5517(-01)$     \\
Percent error in $\alpha_\mathrm{f}$, given that $A_\mathrm{f} = 7.81(-01)$       
& $3.31(+00)$               \\   
\hline

d -- Shell No                         & 13               \\
Inner radius, $\, \xi_{12}$              & $5.0896(+00)$     \\
Outer radius, $\xi_{13}$              & $9.5598(-01)$     \\
Orbit radius, $\, \alpha_\mathrm{d}=\alpha_{13}$            & $5.7711(+00)$     \\
Percent error in $\alpha_\mathrm{d}$, given that $A_\mathrm{d} = 5.76(+00)$       
& $1.93(-01)$               \\   
\hline

\end{tabular}
\end{center}
\end{table}

\begin{table}
\begin{center}
\caption{The 55 Cnc system: shells, in which planets have not been observed. $P_i$ denote predicted orbit radii of planets not yet observed, according to Table~2 of \citep{BL13}; external shells, i.e. shells next to the last occupied shell, are included only when such predictions are available. Percent differences in the computed orbit radii $\alpha$ are given with respect to the corresponding predicted radii $P$, $100 \times |(P - \alpha)| / P$. Other details as in Table \ref{55Cnc}.\label{55CncIE}}
\begin{tabular}{lr} 
\hline \hline
Shell No                                & 3                 \\
Inner radius, $\, \xi_2$                   & $2.5224(-02)$     \\
Outer radius, $\xi_3$                   & $7.7733(-02)$     \\ 
Orbit radius, $\, \alpha_3$                       & $4.4352(-02)$     \\
Percent difference in $\alpha_3$, given that $P_3 = 4.0(-02)$   
& $1.09(+01)$              \\
\hline

Shell No                                & 6                 \\
Inner radius, $\, \xi_5$                   & $3.4540(-01)$     \\
Outer radius, $\xi_6$                   & $5.9715(-01)$     \\
Orbit radius, $\, \alpha_6$                       & $4.5417(-01)$     \\
\hline

Shell No                                & 8                 \\
Inner radius, $\, \xi_7$                   & $9.5598(-01)$     \\
Outer radius, $\xi_8$                   & $1.4396(+00)$     \\
Orbit radius, $\, \alpha_8$                       & $1.1697(+00)$     \\
\hline

Shell No                                & 9                 \\
Outer radius, $\xi_9$                   & $2.0742(+00)$     \\
Orbit radius, $\, \alpha_9$                       & $1.7262(+00)$     \\
Percent difference in $\alpha_9$, given that $P_9 = 1.98(+00)$ & $1.28(+01)$ \\   
\hline

Shell No                                & 10                \\
Outer radius, $\xi_{10}$                & $2.8762(+00)$     \\
Orbit radius, $\, \alpha_{10}$                    & $2.4429(+00)$     \\
\hline

Shell No                                & 11                 \\
Outer radius, $\xi_{11}$                & $3.8765(+00)$     \\
Orbit radius, $\, \alpha_{11}$                    & $3.3377(+00)$     \\
\hline

Shell No                                & 12                \\
Outer radius, $\xi_{12}$                & $5.0896(+00)$     \\
Orbit radius, $\, \alpha_{12}$                    & $4.4358(+00)$     \\
\hline

External Shell No                                & 17       \\
Inner radius, $\, \xi_{16}$                & $1.2621(+01)$     \\
Outer radius, $\xi_{17}$                & $1.5306(+01)$     \\
Orbit radius, $\, \alpha_{17}$                    & $1.3895(+01)$     \\
Percent difference in $\alpha_{17}$, given that $P_{17} = 1.397(+01)$ & $5.41(-01)$ \\
\hline

\end{tabular}
\end{center}
\end{table}

\begin{table}
\begin{center}
\caption{The $\upsilon$ And system: central body $S_1$, i.e. the host star $\upsilon$ And, and polytropic spherical shells of the planets b, c, d, e. Other details as in Table \ref{55Cnc}.\label{uAn}}
\begin{tabular}{lr} 
\hline \hline
Host star $\upsilon$ And -- Shell No                    & 1               \\
$n_\mathrm{opt}$                                        & $2.9490(+00)$   \\
$\xi_1$ (cpu)                                           & $6.7032(+00)$   \\
$\xi_1$ ($R_\odot$)                                     & $1.6310(+00)$   \\
\hline

b -- Shell No                                & 3                 \\
Inner radius, $\, \xi_2$                        & $3.7970(-02)$     \\
Outer radius, $\xi_3$                        & $1.0501(-01)$     \\
Orbit radius, $\, \alpha_\mathrm{b}=\alpha_3$                   & $6.3391(-02)$     \\
Percent error in $\alpha_\mathrm{b}$, given that $A_\mathrm{b} = 5.9(-02)$   
& $7.44(+00)$              \\ 
\hline

c -- Shell No                                           & 8               \\
Inner radius, $\, \xi_7$                                   & $7.2472(-01)$   \\
Outer radius, $\xi_8$                                   & $1.0677(+00)$     \\
Orbit radius, $\, \alpha_\mathrm{c}=\alpha_8$              & $8.5804(-01)$     \\
Percent error in $\alpha_\mathrm{c}$, given that $A_\mathrm{c} = 8.61(-01)$     
& $3.43(-01)$     \\ 
\hline

d -- Shell No                                                 & 11               \\
Inner radius, $\, \xi_{10}$              & $2.1062(+00)$     \\
Outer radius, $\xi_{11}$              & $2.8127(+00)$     \\
Orbit radius, $\, \alpha_\mathrm{d}=\alpha_{11}$            & $2.4360(+00)$     \\
Percent error in $\alpha_\mathrm{d}$, given that $A_\mathrm{d} = 2.55(+00)$       
& $4.47(+00)$               \\   
\hline

e -- Shell No                         & 14                \\
Inner radius, $\, \xi_{13}$              & $4.6458(+00)$     \\
Outer radius, $\xi_{14}$              & $5.7171(+00)$     \\
Orbit radius, $\, \alpha_\mathrm{e}=\alpha_{14}$            & $5.2581(+00)$     \\
Percent error in $\alpha_\mathrm{e}$, given that $A_\mathrm{e} = 5.2456(+00)$       
& $2.38(-01)$               \\   
\hline

\end{tabular}
\end{center}
\end{table}

\begin{table}
\begin{center}
\caption{The $\upsilon$ And system: shells, in which planets have not been observed. $P_i$ denote predicted orbit radii of planets not yet observed, according to Table~2 of \citep{BL13}. Other details as in Table \ref{55CncIE}.\label{uAnIE}}
\begin{tabular}{lr} 
\hline \hline
Shell No                                & 2                 \\
Inner radius, $\, \xi_1$                & $7.5879(-03)$     \\
Outer radius, $\xi_2$                   & $3.7970(-02)$     \\ 
Orbit radius, $\, \alpha_2$             & $1.6980(-02)$     \\
\hline

Shell No                                & 4                 \\
Inner radius, $\, \xi_3$                & $1.0501(-01)$     \\
Outer radius, $\xi_4$                   & $2.2122(-01)$     \\
Orbit radius, $\, \alpha_4$             & $1.5621(-01)$     \\
Percent difference in $\alpha_4$, given that $P_4 = 1.4(-01)$ & $1.16(+01)$ \\             
\hline

Shell No                                & 5                 \\
Outer radius, $\xi_5$                   & $3.8912(-01)$     \\
Orbit radius, $\, \alpha_5$             & $3.1016(-01)$     \\
Percent difference in $\alpha_5$, given that $P_5 = 3.6(-01)$ & $1.38(+01)$ \\             
\hline

Shell No                                & 6                 \\
Outer radius, $\xi_6$                   & $5.3931(-01)$     \\
Orbit radius, $\, \alpha_6$             & $4.6122(-01)$     \\
\hline

Shell No                                & 7                  \\
Outer radius, $\xi_7$                   & $7.2472(-01)$     \\
Orbit radius, $\, \alpha_7$             & $6.3001(-01)$     \\
\hline

Shell No                                & 9                 \\
Inner radius, $\, \xi_8$                & $1.0677(+00)$     \\
Outer radius, $\xi_9$                   & $1.5280(+00)$     \\
Orbit radius, $\, \alpha_9$             & $1.2701(+00)$     \\
\hline

Shell No                                & 10                \\
Outer radius, $\xi_{10}$                & $2.1062(+00)$     \\
Orbit radius, $\, \alpha_{10}$                    & $1.7919(+00)$     \\
\hline

Shell No                                & 12                \\
Inner radius, $\, \xi_{11}$             & $2.8127(+00)$     \\
Outer radius, $\xi_{12}$                & $3.6577(+00)$     \\
Orbit radius, $\, \alpha_{12}$          & $3.2182(+00)$     \\
\hline

External Shell No                       & 20       \\
Inner radius, $\, \xi_{19}$             & $1.2565(+01)$     \\
Outer radius, $\xi_{20}$                & $1.4727(+01)$     \\
Orbit radius, $\, \alpha_{20}$          & $1.3652(+01)$     \\
Percent difference in $\alpha_{20}$, given that $P_{20} = 1.357(+01)$ & $6.04(-01)$ \\
\hline

\end{tabular}
\end{center}
\end{table}

\begin{table}
\begin{center}
\caption{The Kepler-11 system: central body $S_1$, i.e. the host star Kepler-11, and  polytropic spherical shells of the planets b, c, d, e, f, and g. For a shell $S_j$, $\alpha_{Lj}$ and $\alpha_{Rj}$ are average-density orbit radii inside $S_j$, located to the left and to the right of the (maximum-density) orbit radius $\alpha_j$, respectively. Other details as in Table \ref{55Cnc}.\label{k11}}
\begin{tabular}{lr} 
\hline \hline
Host star Kepler-11 -- Shell No                         & 1               \\
$n_\mathrm{opt}$                                        & $2.7790(+00)$   \\
$\xi_1$ (cpu)                                           & $6.1250(+00)$   \\
$\xi_1$ ($R_\odot$)                                     & $1.0650(+00)$   \\
\hline

b -- Shell No                                & 5                 \\
Inner radius, $\, \xi_4$                     & $7.4097(-02)$     \\
Outer radius, $\xi_5$                        & $1.3069(-01)$     \\
Orbit radius, $\, \alpha_\mathrm{b}=\alpha_5$   & $9.3835(-02)$  \\
Percent error in $\alpha_\mathrm{b}$, given that $A_\mathrm{b} = 9.1(-02)$   
& $3.12(+00)$              \\ 
\hline

c -- Shell No                                            & 5               \\
Orbit radius, $\, \alpha_\mathrm{c}=\alpha_\mathrm{R5}$  & $1.1285(-01)$   \\
Percent error in $\alpha_\mathrm{c}$, given that $A_\mathrm{c} = 1.06(-01)$     
& $6.46(+00) $     \\ 
\hline

d -- Shell No                                                 & 6               \\
Outer radius, $\xi_6$                                         & $1.9686(-01)$   \\
Orbit radius, $\, \alpha_\mathrm{d}=\alpha_{L6}$              & $1.5552(-01)$   \\
Percent error in $\alpha_\mathrm{d}$, given that $A_\mathrm{d} = 1.59(-01)$       
& $2.19(+00)$               \\   
\hline

e -- Shell No                                            & 6               \\
Orbit radius, $\, \alpha_\mathrm{e}=\alpha_6$            & $1.7715(-01)$     \\
Percent error in $\alpha_\mathrm{e}$, given that $A_\mathrm{e} = 1.94(-01)$       
& $8.68(+00)$               \\   
\hline

f -- Shell No                         & 7                \\
Outer radius, $\xi_7$                 & $2.6693(-01)$     \\
Orbit radius, $\, \alpha_\mathrm{f}=\alpha_7$     & $2.2517(-01)$     \\
Percent error in $\alpha_\mathrm{f}$, given that $A_\mathrm{f} = 2.5(-01)$       
& $9.93(+00)$               \\   
\hline

g -- Shell No                         & 9                 \\
Inner radius, $\, \xi_8$              & $3.7388(-01)$     \\
Outer radius, $\xi_9$                 & $4.7787(-01)$     \\
Orbit radius, $\, \alpha_\mathrm{g}=\alpha_9$     & $4.5138(-01)$     \\
Percent error in $\alpha_\mathrm{g}$, given that $A_\mathrm{g} = 4.62(-01)$       
& $2.30(+00)$               \\   
\hline

\end{tabular}
\end{center}
\end{table}

\begin{table}
\begin{center}
\caption{The Kepler-11 system: shells, in which planets have not been observed. $P_i$ denote predicted orbit radii of planets not yet observed, according to Table~2 of \citep{BL13}. Other details as in Table \ref{55CncIE}.\label{k11IE}}
\begin{tabular}{lr} 
\hline \hline
Shell No                                & 2                 \\
Inner radius, $\, \xi_1$                & $4.9547(-03)$     \\
Outer radius, $\xi_2$                   & $2.1449(-02)$     \\ 
Orbit radius, $\, \alpha_2$             & $1.0516(-02)$     \\
\hline

Shell No                                & 3                 \\
Outer radius, $\xi_3$                   & $4.6115((-02)$    \\ 
Orbit radius, $\, \alpha_3$             & $3.8828(-02)$     \\
\hline

Shell No                                & 4                 \\
Outer radius, $\xi_4$                   & $7.4097(-02)$     \\
Orbit radius, $\, \alpha_4$             & $5.2580(-02)$     \\
\hline

Shell No                                & 8                 \\
Inner radius, $\, \xi_7$                & $2.6693(-01)$     \\
Outer radius, $\xi_8$                   & $3.7388(-01)$     \\
Orbit radius, $\, \alpha_8$             & $3.1305(-01)$     \\
Percent difference in $\alpha_8$, given that $P_8 = 3.4(-01)$ & $8.58(+00)$ \\             
\hline

External Shell No                       & 20       \\
Inner radius, $\, \xi_{19}$             & $3.2942(+00)$     \\
Outer radius, $\xi_{20}$                & $3.7137(+00)$     \\
Orbit radius, $\, \alpha_{20}$          & $3.6531(+00)$     \\
Percent difference in $\alpha_{20}$, given that $P_{20} = 3.5(+00)$ & $4.37(+00)$ \\
\hline

\end{tabular}
\end{center}
\end{table}

\begin{table}
\begin{center}
\caption{The Kepler-20 system: central body $S_1$, i.e. the host star Kepler-20, and  polytropic spherical shells of the planets b, e, c, f, and d. Other details as in Tables \ref{55Cnc} and \ref{k11}.\label{k20}}
\begin{tabular}{lr} 
\hline \hline
Host star Kepler-20 -- Shell No                    & 1               \\
$n_\mathrm{opt}$                                   & $2.7990(+00)$   \\
$\xi_1$ (cpu)                                      & $6.1881(+00)$   \\
$\xi_1$ ($R_\odot$)                                & $9.4400(-01)$   \\
\hline

b -- Shell No                                & 4                 \\
Inner radius, $\, \xi_3$                     & $4.3317(-02)$     \\
Outer radius, $\xi_4$                        & $6.8027(-02)$     \\
Orbit radius, $\, \alpha_\mathrm{b}=\alpha_\mathrm{L4}$  & $4.5598(-02)$     \\
Percent error in $\alpha_\mathrm{b}$, given that $A_\mathrm{b} = 4.537(-02)$   
& $5.03(-01)$              \\ 
\hline

e -- Shell No                                   & 4               \\
Orbit radius, $\, \alpha_\mathrm{e}=\alpha_4$   & $5.1099(-02)$   \\
Percent error in $\alpha_\mathrm{e}$, given that $A_\mathrm{e} = 5.07(-02)$     
& $7.87(-01)$     \\ 
\hline

c -- Shell No                      & 5               \\
Outer radius, $\xi_5$              & $1.1892(-01)$     \\
Orbit radius, $\, \alpha_\mathrm{c}=\alpha_5$  & $8.3745(-02)$     \\
Percent error in $\alpha_\mathrm{c}$, given that $A_\mathrm{c} = 9.3(-02)$       
& $9.95(+00)$               \\   
\hline

f -- Shell No                      & 5                \\
Orbit radius, $\, \alpha_\mathrm{f}=\alpha_\mathrm{R5}$     & $1.0078(-01)$     \\
Percent error in $\alpha_\mathrm{f}$, given that $A_\mathrm{f} = 1.1(-01)$       
& $8.18(+00)$               \\   
\hline

d -- Shell No                         & 8                \\
Inner radius, $\, \xi_7$              & $2.9372(-01)$     \\
Outer radius, $\xi_8$                 & $3.7114(-01)$     \\
Orbit radius, $\, \alpha_\mathrm{d}=\alpha_8$     & $3.4492(-01)$     \\
Percent error in $\alpha_\mathrm{d}$, given that $A_\mathrm{d} = 3.453(-01)$       
& $1.11(-01)$               \\   
\hline

\end{tabular}
\end{center}
\end{table}

\begin{table}
\begin{center}
\caption{The Kepler-20 system: shells, in which planets have not been observed. $P_i$ denote predicted orbit radii of planets not yet observed, according to Table~2 of \citep{BL13}. Other details as in Table \ref{55CncIE}.\label{k20IE}}
\begin{tabular}{lr} 
\hline \hline
Shell No                                & 2                 \\
Inner radius, $\, \xi_1$                & $4.3917(-03)$     \\
Outer radius, $\xi_2$                   & $1.9359(-02)$     \\ 
Orbit radius, $\, \alpha_2$             & $9.2262(-03)$     \\
\hline

Shell No                                & 3                 \\
Outer radius, $\xi_3$                   & $4.3317(-02)$     \\ 
Orbit radius, $\, \alpha_3$             & $3.4776(-02)$     \\
\hline

Shell No                                & 6                 \\
Inner radius, $\, \xi_5$                & $1.1892(-01)$     \\
Outer radius, $\xi_6$                   & $1.9811(-01)$     \\
Orbit radius, $\, \alpha_6$             & $1.5330(-01)$     \\
\hline

Shell No                                & 7                 \\
Outer radius, $\xi_7$                   & $2.9372(-01)$     \\
Orbit radius, $\, \alpha_7$             & $2.5195(-01)$     \\
Percent difference in $\alpha_7$, given that $P_7 = 2.2(-01)$ & $1.45(+01)$ \\             
\hline

External Shell No                       & 10                \\
Inner radius, $\, \xi_9$                & $4.8901(-01)$     \\
Outer radius, $\xi_{10}$                & $6.6766(-01)$     \\
Orbit radius, $\, \alpha_{10}$          & $5.6847(-01)$     \\
Percent difference in $\alpha_{10}$, given that $P_{10} = 4.9(-01)$ & $1.60(+01)$ \\
\hline

\end{tabular}
\end{center}
\end{table}

\begin{table}
\begin{center}
\caption{The Kepler-90 system: central body $S_1$, i.e. the host star Kepler-90, and  polytropic spherical shells of the planets b, c, d, e, f, g, h. Other details as in Tables \ref{55Cnc} and \ref{k11}.\label{k90}}
\begin{tabular}{lr} 
\hline \hline
Host star Kepler-90 -- Shell No                    & 1               \\
$n_\mathrm{opt}$                                   & $2.8190(+00)$   \\
$\xi_1$ (cpu)                                      & $6.2525(+00)$   \\
$\xi_1$ ($R_\odot$)                                & $1.2000(+00)$   \\
\hline

b -- Shell No                                & 4                 \\
Inner radius, $\, \xi_3$                     & $5.8642(-02)$     \\
Outer radius, $\xi_4$                        & $9.1111(-02)$     \\
Orbit radius, $\, \alpha_\mathrm{b}=\alpha_\mathrm{4}$  & $7.4109(-02)$     \\
Percent error in $\alpha_\mathrm{b}$, given that $A_\mathrm{b} = 7.4(-02)$   
& $1.48(-01)$              \\ 
\hline

c -- Shell No                                            & 4               \\
Orbit radius, $\, \alpha_\mathrm{c}=\alpha_\mathrm{R4}$  & $8.5717(-02)$   \\
Percent error in $\alpha_\mathrm{c}$, given that $A_\mathrm{c} = 8.9(-02)$       
& $3.69(+00)$               \\   
\hline

d -- Shell No                         & 7                 \\
Inner radius, $\, \xi_6$              & $2.6036(-01)$     \\
Outer radius, $\xi_7$                 & $4.0383(-01)$     \\
Orbit radius, $\, \alpha_\mathrm{d}=\alpha_7$     & $3.2412(-01)$     \\
Percent error in $\alpha_\mathrm{d}$, given that $A_\mathrm{d} = 3.2(-01)$       
& $1.29(+00)$               \\   
\hline

e -- Shell No                         & 8                 \\
Outer radius, $\xi_8$                 & $5.8605(-01)$     \\
Orbit radius, $\, \alpha_\mathrm{e}=\alpha_\mathrm{L8}$   & $4.4108(-01)$     \\
Percent error in $\alpha_\mathrm{e}$, given that $A_\mathrm{e} = 4.2(-01)$     
& $5.02(+00)$     \\ 
\hline

f -- Shell No                                     & 8                 \\
Orbit radius, $\, \alpha_\mathrm{f}=\alpha_8$     & $4.9108(-01)$     \\
Percent error in $\alpha_\mathrm{f}$, given that $A_\mathrm{f} = 4.8(-01)$       
& $2.31(+00)$               \\   
\hline

g -- Shell No                         & 9                 \\
Outer radius, $\xi_9$                 & $7.6805(-01)$     \\
Orbit radius, $\, \alpha_\mathrm{g}=\alpha_9$   & $7.1520(-01)$     \\
Percent error in $\alpha_\mathrm{g}$, given that $A_\mathrm{g} = 7.1(-01)$     
& $7.32(-01)$     \\ 
\hline

h -- Shell No                                & 11                \\
Inner radius, $\, \xi_{10}$                  & $9.1537(-01)$     \\
Outer radius, $\xi_{11}$                     & $1.1800(+00)$     \\
Orbit radius, $\, \alpha_\mathrm{h}=\alpha_{11}$  & $1.0081(+00)$     \\
Percent error in $\alpha_\mathrm{h}$, given that $A_\mathrm{h} = 1.01(+00)$   
& $1.92(-01)$              \\ 
\hline

\end{tabular}
\end{center}
\end{table}

\begin{table}
\begin{center}
\caption{The Kepler-90 system: internal shells, in which planets have not been observed, and the lower three external shells. Other details as in Table \ref{55CncIE}.\label{k90IE}}
\begin{tabular}{lr} 
\hline \hline
Shell No                                & 2                 \\
Inner radius, $\, \xi_1$                & $5.5827(-03)$     \\
Outer radius, $\xi_2$                   & $2.5047(-02)$     \\ 
Orbit radius, $\, \alpha_2$             & $1.1608(-02)$     \\
\hline

Shell No                                & 3                 \\
Outer radius, $\xi_3$                   & $5.8642(-02)$     \\ 
Orbit radius, $\, \alpha_3$             & $4.4644(-02)$     \\
\hline

Shell No                                & 5                 \\
Inner radius, $\, \xi_4$                & $9.1111(-02)$     \\
Outer radius, $\xi_5$                   & $1.5507(-01)$     \\
Orbit radius, $\, \alpha_5$             & $1.0715(-01)$     \\
\hline

Shell No                                & 6                 \\
Outer radius, $\xi_6$                   & $2.6036(-01)$     \\
Orbit radius, $\, \alpha_7$             & $1.9911(-01)$     \\
\hline

Shell No                                & 10       \\
Inner radius, $\, \xi_9$                & $7.6805(-01)$     \\
Outer radius, $\xi_{10}$                & $9.1537(-01)$     \\
Orbit radius, $\, \alpha_{10}$          & $8.1699(-01)$     \\
\hline

External Shell No                       & 11                \\
Outer radius, $\xi_{11}$                & $1.1800(+00)$     \\
Orbit radius, $\, \alpha_{11}$          & $1.0081(+00)$     \\
\hline

External Shell No                       & 12                \\
Outer radius, $\xi_{12}$                & $1.5231(+00)$     \\
Orbit radius, $\, \alpha_{12}$          & $1.3384(+00)$     \\
\hline

External Shell No                       & 13                \\
Outer radius, $\xi_{13}$                & $1.8991(+00)$     \\
Orbit radius, $\, \alpha_{13}$          & $1.7286(+00)$     \\
\hline

\end{tabular}
\end{center}
\end{table}

\end{document}